\documentclass[conference]{IEEEtran}
\IEEEoverridecommandlockouts
\usepackage{enumitem}
\usepackage{bm}
\usepackage{cite}
\usepackage{hyperref}
\usepackage{amsmath}
\usepackage{booktabs}
\usepackage{afterpage}
\usepackage{amsmath,amssymb,amsfonts}
\usepackage{algorithmic}
\usepackage{graphicx}
\usepackage{textcomp}
\usepackage{xcolor}
\usepackage{subcaption}
\def\BibTeX{{\rm B\kern-.05em{\sc i\kern-.025em b}\kern-.08em
    T\kern-.1667em\lower.7ex\hbox{E}\kern-.125emX}}
\begin{document}

\title{UAV SAR Imaging with 5G NR OFDM Signals in NLOS Environments\\

}
\author{Qiuyuan~Yang, Cunhua~Pan, Ruidong~Li, Zhenkun~Zhang, Hong~Ren, Changhong~Wang, \\
Jiangzhou~Wang,~\IEEEmembership{Fellow,~IEEE}\\
{\footnotesize \textsuperscript{}}

\thanks{Qiuyuan~Yang, Cunhua~Pan, Zhenkun~Zhang, Hong~Ren and Jiangzhou~Wang are with National Mobile Communications Research Laboratory, Southeast University, Nanjing 210096, China (email: 220241246, cpan, zhenkun\_zhang, hren, j.z.wang@seu.edu.cn).

Ruidong~Li, Changhong~Wang are with Jinan Maiwei Intelligent Technology Co., Ltd (email: lird, wangchh01@inspur.com).
}
}

\maketitle

\begin{abstract}
The integration of sensing and communication (ISAC) has significant potential for future wireless systems, enabling efficient spectrum utilization and novel application scenarios. In this paper, we propose a cooperative ISAC framework for synthetic aperture radar (SAR) imaging by leveraging orthogonal frequency division multiplexing (OFDM) communication signals. We address the challenge of severe imaging degradation in non-line-of-sight (NLOS) environments under the QUAsi Deterministic RadIo channel GenerAtor (QuaDRiGa). To detect weak signals and eliminate false points, we develop a two-stage compressed sensing-space alternating generalized expectation maximization (CS-SAGE) scheme for high-precision scatterer localization. In stage I, orthogonal matching pursuit (OMP) is employed for coarse estimation to identify the approximate locations of dominant scatterers. Then, the SAGE algorithm in stage II performs fine estimation to accurately extract scatterer parameters. Simulation results validate the effectiveness of the proposed cooperative ISAC framework, and provide valuable insights for practical system design.
\end{abstract}

\begin{IEEEkeywords}
Integrated sensing and communication (ISAC), synthetic aperture radar (SAR) imaging, OMP, SAGE
\end{IEEEkeywords}

\section{Introduction}
With the vision of sixth-generation (6G) mobile communication unfolding, intelligent devices equipped with high-capacity communication and advanced sensing capabilities have drawn much attention. Integrated sensing and communication (ISAC) has been envisioned as a pivotal technology for future networks \cite{wang2021symbiotic}\cite{liu2020joint}. By sharing hardware platforms, spectrum resources, and signal processing mechanisms, ISAC deeply integrates conventional radar sensing and cellular communication. This integration not only reduces the cost and energy consumption of the system, but also significantly improves the efficiency of the spectrum, enabling disruptive applications such as vehicular networks, smart cities, and industrial internet of things (IoT) \cite{zhu2024enabling}\cite{meng2023uav}.

Meanwhile, the low-altitude economy is developing rapidly, with the unmanned aerial vehicles (UAVs) serving as its main carriers \cite{lu2022degrees}. 
Synthetic aperture radar (SAR) is commonly used for UAV observation because of its all-time, all-weather operability and high-resolution imaging capability \cite{koo2012new}. Unlike optical sensors that are sensitive to illumination and weather conditions, SAR can penetrate clouds, fog, and vegetation, to produce clear images of the environment. As radar signals differ from communication signals, achieving simultaneous sensing and communication capabilities necessitates installing both radar and communication equipment on the UAV. This not only increases costs and energy consumption but also introduces delays and Doppler inconsistencies during the post-processing fusion of sensing and communication data, leading to mutual interference between systems. Joint communication and sensing on UAV platforms is therefore an attractive solution to support low-altitude connectivity and perception simultaneously \cite{zhang2024target}\cite{tang2025cooperative}.

Recent studies have explored leveraging communication signals for radar sensing and imaging. The authors of \cite{zhang2014ofdm} employed CP-OFDM signals as the transmission signals for SAR radar and drawn the conclusion that sufficiently long cyclic prefixes can eliminate intersymbol interference. In \cite{zhang2019joint}, the waveforms for communication and sensing were separated in the time-frequency spectrum, demonstrating that OFDM can be utilised to achieve the integration of communication and sensing tasks. The work \cite{zheng2024random} further indicated that altering the modulation scheme of signals can prioritise either perception imaging or communication transmission. Considering the impact of multipath propagation, \cite{moro2024exploring} investigated SAR schemes in realistic scenarios but did not address the potential issues of positional offset and ghost points.

The main contributions of this paper are summarized as follows:

\begin{enumerate}[label=\arabic*)]
    \item We propose a novel ISAC system that leverages 5G NR OFDM signals to achieve SAR imaging. The proposed framework integrates communication and sensing functionalities, enabling high-resolution SAR imaging without requiring dedicated radar waveforms. 
    \item To address the challenges of non-line-of-sight (NLOS) environments, we propose a two-stage compressed sensing–space alternating generalized expectation maximization (CS-SAGE) scheme. This approach effectively mitigates the adverse effects of multipath propagation, while reducing computational complexity and  enhancing imaging precision.
    \item We validate the effectiveness of the proposed cooperative ISAC framework and sensing scheme based on the simulation results. Insights about the performance of ISAC systems are also derived from the simulation results, providing guidance for engineering practice.
\end{enumerate}

\section{SYSTEM MODEL}
\subsection{5G NR SSB}
\begin{figure}[!t]
    \centering
    \includegraphics[width=\columnwidth]{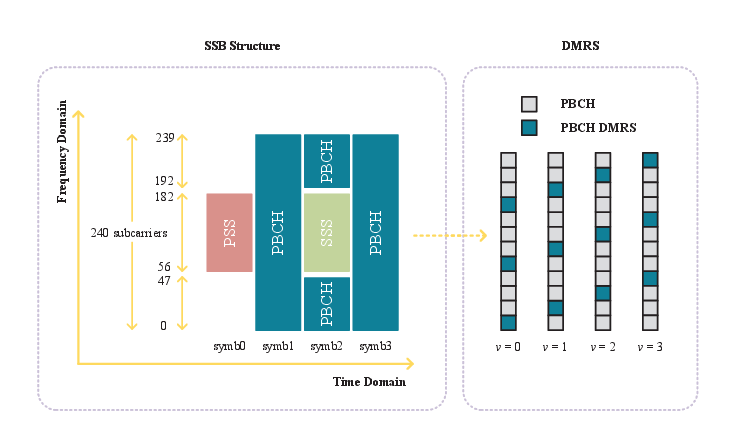}
    \caption{SSB time-frequency domain structure.}
    \label{fig:fig2}
\end{figure}

Synchronization signal block (SSB) is leveraged as a fundamental reference signal, owing to its well-defined structure and periodic transmission. As shown in Fig.~\ref{fig:fig2}, each SSB is a self-contained resource unit occupying 20 resource blocks (RBs) in frequency domain over four consecutive OFDM symbols, where each RB is composed of 12 consecutive subcarriers. The Primary Synchronization Signal (PSS) occupies the central 127 subcarriers of symbol 0. It allows the UAV to achieve initial symbol timing and acquire the physical-layer identity $N_{\mathrm{ID}}^{2}$ within a cell group. The Secondary Synchronization Signal (SSS), found in symbol 2 and also covering 127 subcarriers, provides frame timing and the physical-layer cell-group identity $N_{\mathrm{ID}}^{1}$. Together, the PSS and SSS determine the full Physical Cell ID (PCI). The Physical Broadcast Channel (PBCH) is transmitted in symbols 1 and 3, as well as the parts of symbol 2 not occupied by the SSS. It carries the Master Information Block (MIB), which
contains the most critical parameters for initial cell access. 

To ensure reliable decoding of the MIB, Demodulation Reference Signals (DMRS) are embedded within the PBCH payload. A key feature of the NR design is that the positions of these DMRS on the subcarrier grid are not static. Instead, they follow a specific pattern that is cyclically shifted in frequency domain according to a parameter $v$, which is derived directly from the cell's PCI . This deliberate placement effectively randomizes interference between the DMRS of neighboring cells, significantly improving PBCH decoding performance, particularly at cell edges. For our sensing application, the known and predictable structure of the DMRS provides a stable and reliable phase reference that can be exploited.

\begin{figure}[!t]
    \centering
    \includegraphics[width=\columnwidth]{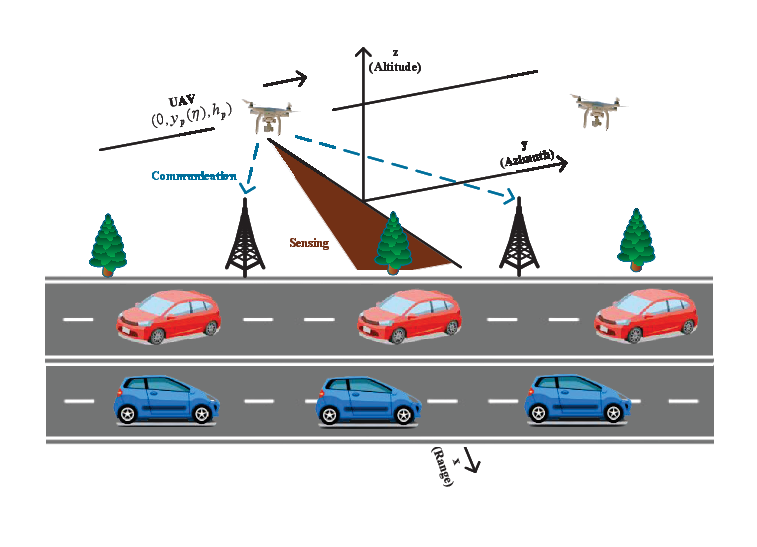}
    \caption{Monostatic OFDM SAR geometry.}
    \label{fig:fig1}
\end{figure}

\subsection{OFDM SAR Receiving Signal Model}
As shown in Fig.~\ref{fig:fig1}, we consider a monostatic OFDM SAR platform carried by a UAV. The UAV periodically transmits OFDM signals while receiving backscattered echoes from ground targets during flight. Without loss of generality, we assume that the UAV flies along the y-axis. The platform's motion enables azimuth-axis synthetic aperture, thereby achieving high-resolution imaging capability. 

Consider an OFDM transmitter with $N$ subcarriers, with the subcarrier spacing, bandwidth, and symbol duration being given as $\Delta f$, $B = N\Delta f$, and $T = 1/\Delta f$. Then, the baseband time-domain OFDM signal within a symbol duration can be represented as
\begin{equation}
s(t) = \frac{1}{\sqrt{N}} \sum_{k=0}^{N-1} S_k \exp\{j2\pi k \Delta f (t - T_{\mathrm{CP}})\}, \ t \in [0, T + T_{\mathrm{CP}}],
\end{equation}
where $S_k$ satisfying $\sum_{k=0}^{N-1} |S_k|^2 = N$ denotes the modulation symbol on the $k$th subcarrier, which ensures that the average symbol power is 1, and $T_{\mathrm{CP}}$ is the time duration of the CP.

The instantaneous position of the UAV can be expressed as $(0,y_p(\eta),h_p)$, where $\eta$ denotes the slow time index in SAR imaging, and $h_p$ represents the altitude of the UAV. The azimuth envelope is defined as
\begin{equation}
\varepsilon_a(\eta)=p^2_a(\theta(\eta)), 
\end{equation}
where $p_a(\theta)=\operatorname{sinc}(0.886\theta/\beta_{bw}),\operatorname{sinc}(x)=\operatorname{sin}(x)/x$ is the sinc function, $\theta$ is the incidence angle of the UAV antenna, $\beta_{bw}=0.886\lambda/L_a$ is the azimuth beamwidth, $L_a$ is the effective length of the antenna. For stationary ground targets, when the UAV is operating in swath mode and illuminates the target scene, the echo signal from the $m$th range cell can be expressed as
\begin{equation}
\begin{split}
u_m(t,\eta) 
&= g_m \, \varepsilon_a(\eta) 
   \exp\!\left(-j \tfrac{4\pi f_{\mathrm{c}} R_m(\eta)}{c}\right) \\
&\quad \times \frac{1}{\sqrt{N}} 
   \sum_{k=0}^{N-1} S_k \,
   \exp\!\left(j \tfrac{2\pi k}{T} \left[t-\tfrac{2R_m(\eta)}{c}\right]\right) \\
&\quad + \, w(t,\eta),t \in [\tfrac{2R_m(\eta)}{c}, \tfrac{2R_m(\eta)}{c}+T + T_{\mathrm{CP}}],
\end{split}
\end{equation}
where $g_m$ is the radar cross-section (RCS), $f_{\mathrm{c}}$ is the carrier frequency, $c$ is the speed of light, $R_m(\eta) = \sqrt{\bar{R}_m^{2} + v_p^{2}\eta^{2}}$ is the instantaneous slant range between the UAV and the $m$th range cell, and $v_p$ is the velocity of the UAV. $\bar{R}_m^{2}$ represents the slant range between the UAV and the $m$th range cell, determined by the range distance and the UAV's flight altitude, and $w(t,\eta)$ represents the noise. By summing over all $M$ range cells, the received signal becomes
\begin{equation}
u(t,\eta) = \sum_{m=0}^{M-1} u_m(t,\eta).
\end{equation}
The system model establishes the received signal in NLOS environment, which forms the foundation for subsequent parameter estimation. Based on this model, the following section focuses on extracting delay and doppler information from the received echoes for accurate target reconstruction.

\section{Algorithm Design}
To address the signal distortion and imaging ambiguities caused by NLOS propagation, a two-stage algorithm for high-precision parameter estimation is proposed in this section. This scheme is designed to effectively undermine multipath effects and ensure the accuracy of the final image.

\subsection{Stage I: OMP-based Coarse Estimation}
In NLOS environment, multipath propagation leads to signal attenuation and the introduction of false target information. The attenuated LOS signal makes it difficult to detect the true target, while the delayed multipath components generate spurious scattering points. To address this issue, we introduce a coarse estimation step based on the OMP algorithm, which effectively reconstructs the signal and suppresses false targets caused by multipath effects. The received signal is expressed as the convolution of the transmitted signal with the impulse response $h(t, \eta)$ as follows
\begin{equation}
u(t,\eta) = s(t) * h(t,\eta) + w(t,\eta).
\end{equation}
Specifically, the channel can be represented as a superposition of multipath effects
\begin{equation}
h(t,\eta) = \sum_{\ell=1}^{L} \alpha_\ell(\eta)\,
\delta\!\big(t-\tau_\ell(\eta)\big)\,
e^{j2\pi f_{d,\ell}(\eta) t},
\end{equation}
where $L$ denotes the number of multipaths, $\alpha_\ell(\eta)$ represents the complex gain of the $\ell$th path, which encompasses both amplitude and phase information, $\tau_\ell(\eta)$ denotes the propagation delay of the $\ell$th path, and $f_{d,\ell}(\eta)$ denotes the Doppler shift of the $\ell$th path. The sensing parameters for each path can be expressed as $\bm\Theta_l = \{\alpha_\ell,\tau_\ell,f_{d,\ell}\}$. From (1), the time-delay Doppler echo model can be derived as follows
\begin{equation}
u(t,\eta) = \sum_{\ell=1}^{L} \alpha_\ell(\eta)\, 
s\!\big(t-\tau_\ell(\eta)\big)\,
e^{j2\pi f_{d,\ell}(\eta) t} + w(t,\eta).
\end{equation}

To formulate the compressed sensing problem, we discretize the continuous parameter space $(\tau, f_d)$ into a grid with delay resolution $\Delta\tau=1/B$ and Doppler resolution $\Delta f_d=1/(N_aT)$, where $N_a$ denotes the number of azimuth samples. For each grid point $(\tau_\ell,f_{d,\ell})$, we construct a steering vector

\begin{align}
\mathbf{a}(\tau_\ell, f_{d,\ell}) 
&= \big[s(t_0 - \tau_\ell)e^{j2\pi f_{d,\ell}t_0}, \ldots, \notag\\
&\quad s(t_{N_t-1} - \tau_\ell)e^{j2\pi f_{d,\ell}t_{N_t-1}}\big]^T,
\end{align}
where $t_n, n=0,...,N_t-1$ are the fast-time sampling instants. Define a dictionary matrix as follows

\begin{equation}
\mathbf{\Phi} = [\mathbf{a}(\tau_0, f_{d,0}), \ldots, \mathbf{a}(\tau_{P-1}, f_{d,Q-1})] \in \mathbb{C}^{N_t \times D},
\end{equation}
where $D=P\times Q$ is the total number of vectors. Vectorizing the received signal gives the sparse model:

\begin{equation}
\mathbf{u} = \mathbf{\Phi} \mathbf{x} + \mathbf{w},
\end{equation}
where $\mathbf{u}$ is the discrete observation vector obtained by sampling the continuous-time received signal $u(t, \eta)$ in the fast-time dimension, $\mathbf{x}\in\mathbb{C}^{D\times 1}$ is the sparse coefficient vector with $||\textbf{x}||_0=L\ll D$. The OMP algorithm iteratively selects atoms that maximize the correlation with the residual signal. At iteration $n$, the correlation is computed as
\begin{equation}
\mathbf{c}^{(n)} = \mathbf{\Phi}^\textnormal{H} \mathbf{r}^{(n-1)},
\end{equation}
and the best atom index is selected as
\begin{equation}
i_n = \underset{{i \notin \mathcal{S}^{(n-1)}}}{argmax}\left|\mathbf{c}^{(n)}_i\right|.
\end{equation}
The support set is updated as $\mathcal{S}^{(n)}=\mathcal{S}^{(n-1)}\cup \left\{i_n \right\}$, where $\mathcal{S}^{(n)}$ represents the set of indices for all atoms that have been previously identified up to the $n$th iteration. The sparse coefficients are estimated via least squares:
\begin{equation}
\hat{\mathbf{x}}_{\mathcal{S}^{(n)}} = \left(\mathbf{\Phi}_{\mathcal{S}^{(n)}}^H \mathbf{\Phi}_{\mathcal{S}^{(n)}}\right)^{-1} \mathbf{\Phi}_{\mathcal{S}^{(n)}}^H \mathbf{u}.
\end{equation}
Then, we can calculate the residual signal as follows
\begin{equation}
\mathbf{r}^{(n)} = \mathbf{u} - \mathbf{\Phi}_{\mathcal{S}^{(n)}} \hat{\mathbf{x}}_{\mathcal{S}^{(n)}}.
\end{equation}
The iterations terminate when the normalized residual energy $\eta^{(n)}=||\mathbf{r}^{(n)}||^2_2/||\mathbf{u}||^2_2$ falls below the threshold $\eta_{th}=\sigma_n^2/||\mathbf{u}||^2_2$, where $\sigma_n^2$ is the noise power estimated by measuring the receiver output when no signal is transmitted. The output provides coarse estimates $\left\{\hat{\bm\Theta}^{\mathrm{OMP}}_\ell\right\}^L_{\ell=1}$ where $\hat{\bm\Theta}^{\mathrm{OMP}}_\ell=\left\{\hat{\alpha}_\ell,\hat{\tau}_\ell,\hat{f}_{d,\ell}\right\}$, which serve as the initialization for the SAGE algorithm.

\subsection{Stage II: SAGE-based Accurate Estimation}
As a method for channel estimation, SAGE mitigates the effects of delay and Doppler shift caused by multipath propagation. Through iterative updates within an expectation-maximisation framework, it enables the separate estimation of delay, Doppler shift, and complex amplitude for each path. Its core principle involves temporarily fixing the results for other paths during each iteration, thereby focusing solely on updating a single path\cite{yin2016performance}. This approach reduces the dimensionality of parameters and accelerates convergence.

The first step is the E-step. The initial parameter estimates $\bm\Theta^{(0)}=\{\bm\Theta_1^{(0)},\bm\Theta_2^{(0)},...,\bm\Theta_L^{(0)}\}$ are derived from the OMP results. Within the generated time-delay Doppler spectrum, the $L$ strongest energy peaks are identified. The corresponding time delays, Doppler shifts, and complex amplitudes form the initial estimate for $\bm\Theta_\ell$.
When updating the parameters of the $\ell$th path, we subtract the current estimated signals from all other paths from the total received signal, thereby isolating a signal comprising solely the $\ell$th path signal and noise. In the $i$th iteration, the purified signal $u_\ell^{(i)}(t,\eta)$ constructed to estimate the $\ell$th path parameters is
\begin{equation}
u_\ell^{(i)}(t, \eta)=u(t, \eta)-\sum_{x=1,x\neq \ell}^{L}\hat{u}_x^{(i-1)}(t, \eta),
\end{equation}
where $\hat{u}_x^{(i-1)}(t, \eta)=\hat{\alpha}_x^{(i-1)}s(t-\hat{\tau}_x^{(i-1)})e^{j2\pi\hat{f}_{d,x}^{(i-1)}t}$ is the signal of the $x$th path reconstructed from the parameter estimates obtained in the $i-1$ th iteration.

The second step is the M-step. The maximum likelihood estimation updates the parameters of each path to maximise the correlation between $u_\ell^{(i)}(t,\eta)$ and $s(t)$. The time delay and Doppler shift are updated to
\begin{equation}
(\hat{\tau}_\ell^{(i)}, \hat{f}_\ell^{(i)}) = \underset{\tau_\ell, f_\ell}{\arg\max} \left| \int_{-\infty}^{\infty} u_\ell^{(i)}(t, \eta) s^*(t-\tau) e^{-j2\pi f_{d,\ell} t} dt \right|^2.
\end{equation}
The amplitude is updated to
\begin{equation}
\hat{\alpha}_\ell^{(i)} = \frac{\int_{-\infty}^{\infty} u_\ell^{(i)}(t, \eta) \cdot s^*(t - \hat{\tau}_\ell^{(i)}) \cdot e^{-j2\pi\hat{f}_{d,\ell}^{(i)}t} dt}{\int_{-\infty}^{\infty} |s(t)|^2 dt}.
\end{equation}

After updating the parameters for path $\ell$, we repeat the E-Step and M-Step for path $\ell+1$, continuing until all $L$ paths have been updated once, thus completing one iteration. This entire process is repeated until the parameter estimates converge. Following the SAGE iteration, an estimated parameter set is obtained for each path. In NLOS environment, paths involving reflections and obstructions exhibit larger delays. Therefore, to mitigate multipath effects, delay serves as the primary parameter for path selection. We statistically determine the minimum delay $\tau_{min}=\underset{r\in{L}}{min}\hat{\tau}_r$ and set a threshold $\tau_{th}(e.g.,\tau_{th}= 1.2\tau_{min})$.Among paths satisfying $\hat{\tau}_r\leq\tau_{th}$, we select the path $q$ with the maximum complex amplitude
\begin{equation}
|\alpha_q|=\underset{l:\tau_r\leq 1.2\tau_{min}}{argmax}|\hat{\alpha}_r|.
\end{equation}

This selection avoids interference from strong reflection paths. The ultimately selected path is employed as the channel response for SAR imaging.

\section{SIMULATION RESULTS}
\begin{figure}[!t]
    \centering
    \begin{subfigure}[b]{0.48\columnwidth}
        \centering
        \includegraphics[width=\textwidth]{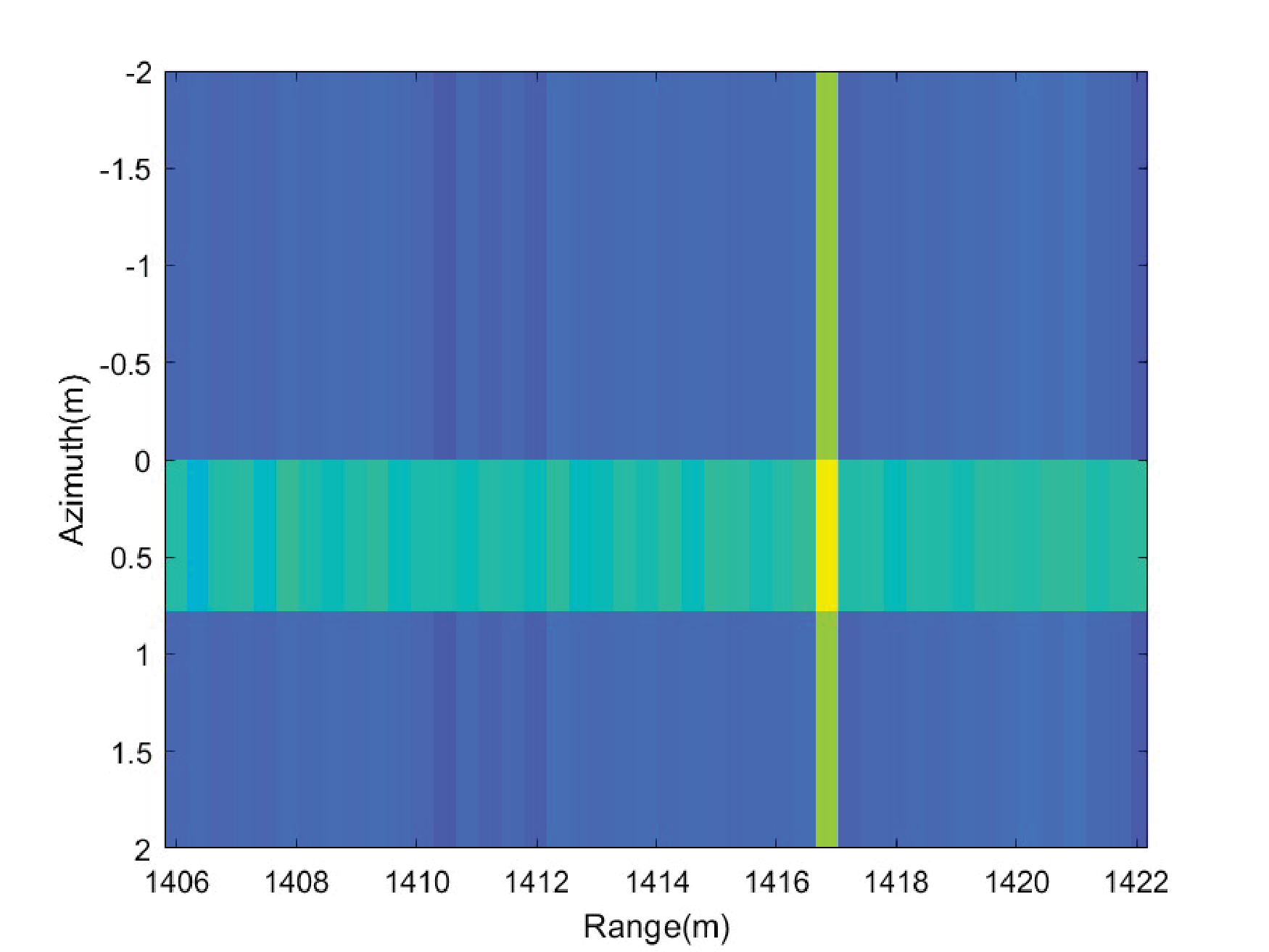}
        \subcaption{}
        \label{point:sub_a}
    \end{subfigure}
    \hfill
    \begin{subfigure}[b]{0.48\columnwidth}
        \centering
        \includegraphics[width=\textwidth]{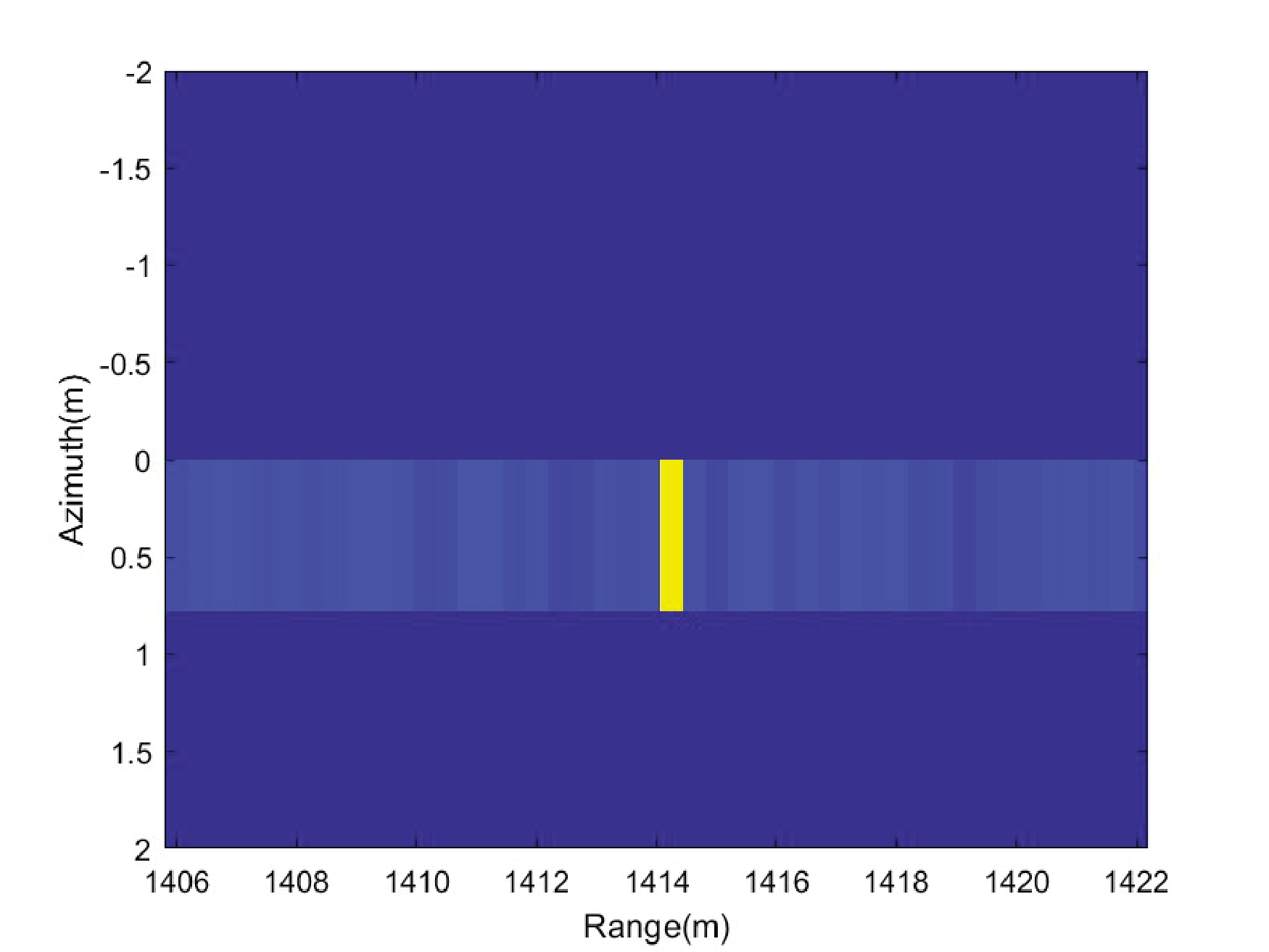}
        \subcaption{}
        \label{point:sub_b}
    \end{subfigure}

    \caption{SAR imaging of point target when SNR is 15dB. (a) Unprocessed imaging results. (b) Imaging results processed by CS-SAGE.}

    \label{fig:point target}
\end{figure}
This section evaluates the performance of the proposed CS-SAGE scheme for the ISAC-SAR framework through simulation experiments. As shown in Table~\ref{tab:params}, the simulation parameters are set following a typical configuration for the 5G NR communications. The propagation channel is generated using the QuaDRiGa simulator under the NTN Rural NLOS scenario to emulate realistic multipath environments. The output signals employ Zero Forcing (ZF) imaging for pulse compression and matched filter imaging for azimuth.
\begin{table}[htbp]
    \caption{Simulation parameters}
    \label{tab:params}
    \centering
    \begin{tabular}{cc}
        \toprule
        \textbf{Parameter} & \textbf{Value} \\
        \midrule
        $f_c$ & 26 GHz \\
        $SCS$ & 120 KHz \\
        $M_{\mathrm{CP}}$ & 36 \\
        $M_{\mathrm{FFT}}$ & 4096 \\
        Data Symbol Duration $T$ & 8.33 $\mu$s \\
        Bandwidth $B$ & 400 MHz \\
        $PRF$ & 800Hz \\
        $h_p$ & 1 KM \\
        $v_p$ & 40 m/s \\
        QuaDriGa NTN Scenario & QuaDRiGa\_NTN\_Rural\_NLOS \\
        QuaDriGa Standard Scenario & 5G-ALLSTAR\_Rural\_LOS \\
        \bottomrule
    \end{tabular}
\end{table}
\subsection{Imaging Results}
\begin{figure}[!t]
    \centering
    \begin{subfigure}[b]{0.48\columnwidth}
        \centering
        \includegraphics[width=\textwidth]{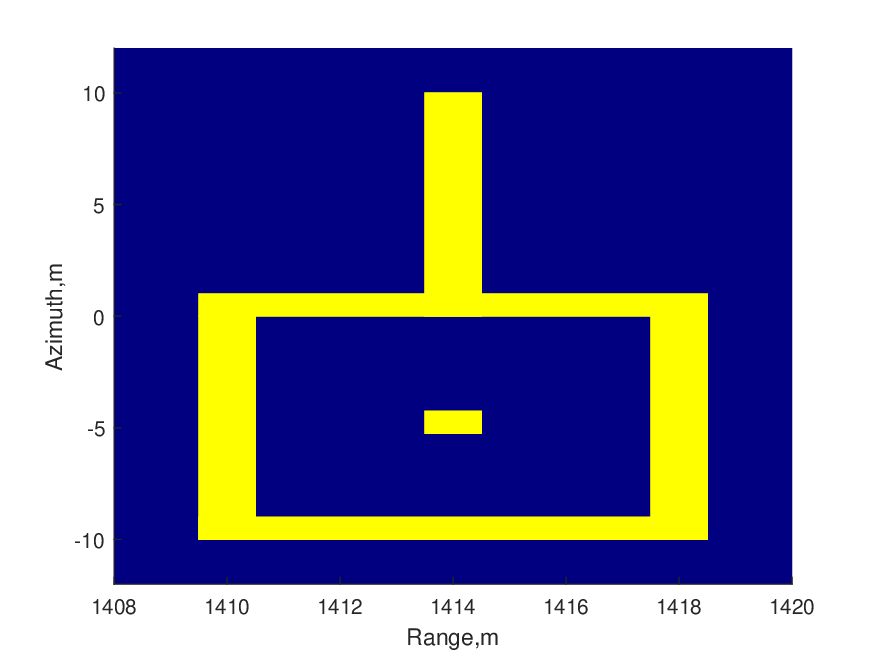}
        \subcaption{}
        \label{extend:sub_a}
    \end{subfigure}
    \hfill
    \begin{subfigure}[b]{0.48\columnwidth}
        \centering
        \includegraphics[width=\textwidth]{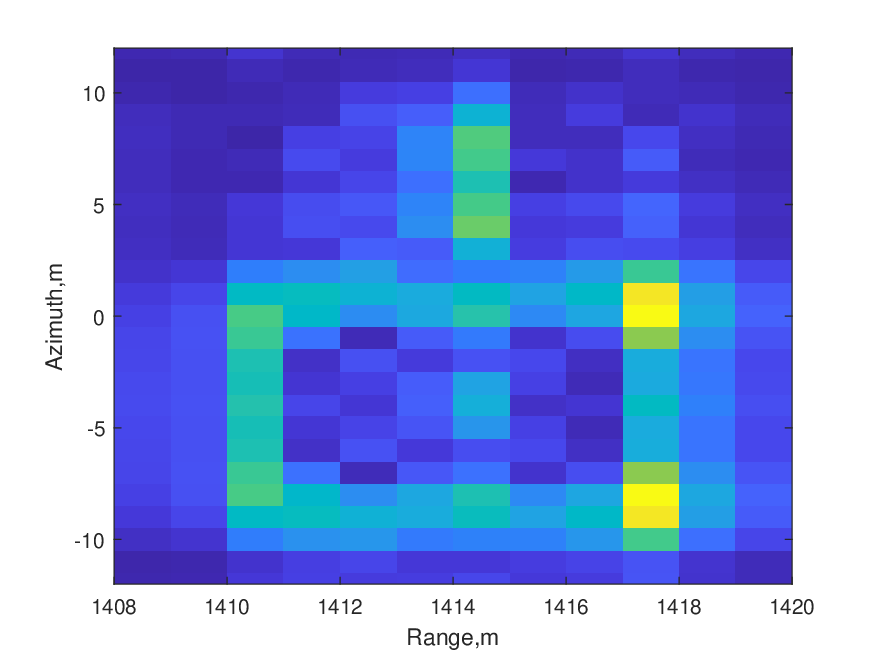}
        \subcaption{}
        \label{extend:sub_b}
    \end{subfigure}

    \caption{SAR imaging of extend target when SNR is 15dB. (a) Origin image. (b) Imaging results processed by CS-SAGE.}

    \label{fig:extended target}
\end{figure}

To validate the effectiveness of the proposed CS-SAGE scheme, we examine SAR imaging under NLOS conditions at SNR = 15 dB. Fig.~\ref{fig:point target} presents imaging results for a single point target, while Fig.~\ref{extend:sub_a} illustrates imaging of an extended target model, with each edge defined by assigning non-zero RCS values to 100 scattering points.

Without CS-SAGE processing, the images in Fig.~\ref{point:sub_a} exhibit severe degradation characteristic of NLOS multipath propagation. The most prominent issue is the range offset from the true target positions. In addition, the imaging energy is diffusely distributed rather than concentrated at the true target locations, and the pronounced sidelobe artifacts spread across the image plane. Furthermore, spurious scattering points appear as ghost images, created by strong multipath reflections.

The CS-SAGE processing in Fig.~\ref{point:sub_b} and Fig.~\ref{extend:sub_b} successfully addresses these issues through a two-stage mechanism that fundamentally differs from conventional SAR processing. In the first stage, OMP exploits the sparse nature of the multipath channel to perform multipath disambiguation. The threshold ensures that the algorithm terminates before including spurious paths, effectively suppressing the ghost artifacts visible in the unprocessed images. Building upon this coarse multipath profile, the second stage employs SAGE to refine the delay and Doppler parameters of each identified path through iterative expectation-maximization. The final path selection strategy identifies the near-direct propagation path by choosing the minimum-delay component with maximum amplitude. It ensures that the imaging uses the true target echo rather than multipath reflections. As a result, the range offset is eliminated through accurate direct path delay estimation, the imaging energy is tightly focused at the correct target locations with significantly suppressed sidelobes, and the spurious artifacts are removed.

\subsection{Performance Analysis}
\begin{figure}[h]
    \centering
    \begin{subfigure}[b]{0.48\columnwidth}
        \centering
        \includegraphics[width=\textwidth]{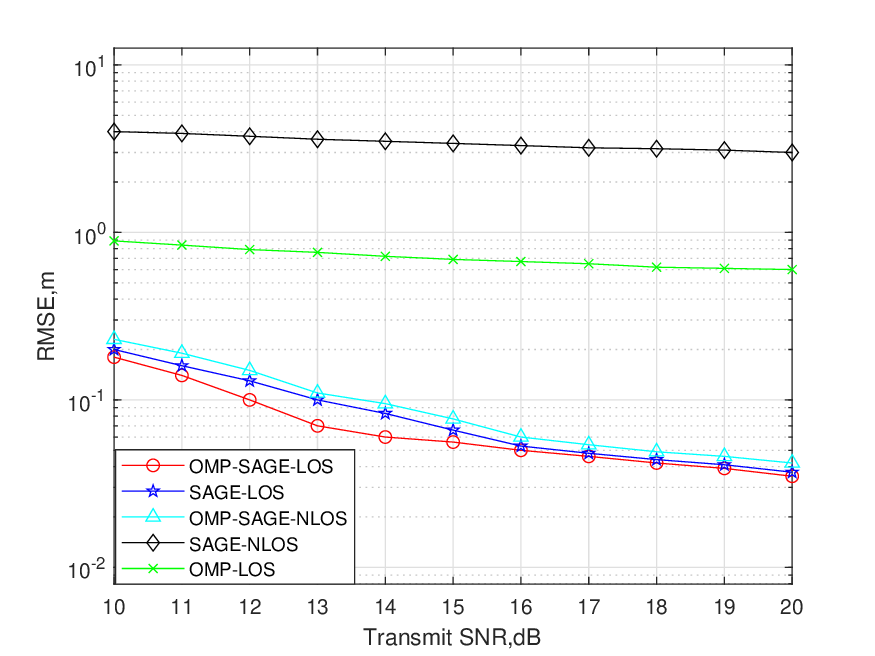}
        \subcaption{}
        \label{MSE:sub_a}
    \end{subfigure}
    \hfill
    \begin{subfigure}[b]{0.48\columnwidth}
        \centering
        \includegraphics[width=\textwidth]{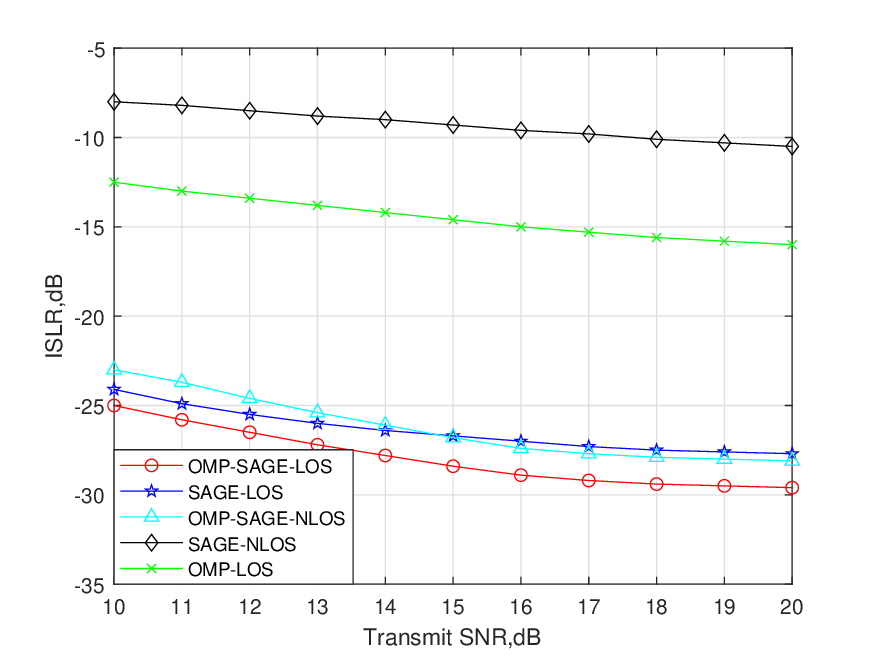}
        \subcaption{}
        \label{ISLR:sub_b}
    \end{subfigure}

    \caption{Performance comparison between different signals. (a) Estimation RMSE vs.SNR. (b) Estimation ISLR vs.SNR.}

    \label{fig:Performance Analysis}
\end{figure}

To evaluate the proposed CS-SAGE scheme, we compare its performance with algorithm across varying SNR conditions ranging from 10 to 20 dB. Fig.~\ref{MSE:sub_a} presents the RMSE of target localization, while Fig.~\ref{ISLR:sub_b} shows the ISLR as a measure of imaging quality. 

In the LOS scenario, both the SAGE and OMP-SAGE methods achieve nearly identical performance. This performance parity occurs because the LOS channel contains a dominant direct path with minimal multipath interference, allowing SAGE's iterative expectation-maximization to reliably converge to the correct parameters even without careful initialization. In this benign propagation environment, the OMP preprocessing stage provides negligible additional benefit since there are no strong multipath ambiguities to resolve. 

However, the NLOS environment reveals a dramatically different picture where the two-stage architecture becomes essential. The SAGE-only method suffers severe performance degradation even at high SNR. This failure occurs because NLOS propagation creates strong multipath components with similar received powers, transforming the parameter estimation problem into a optimization filled with local optima. Without proper initialization, SAGE cannot distinguish which peak corresponds to the true direct path versus reflected paths, frequently converging to incorrect multipath components. By contrast, the OMP-SAGE maintains robust performance. OMP separates dominant multipath components from noise and weak reflections, providing SAGE with initialisation close to the true solution to converge to correct direct path parameters.

\section{Conclusion}
In this paper, we proposed a novel framework for achieving high-precision UAV SAR imaging within a 5G NR simulation environment. Imaging and performance analysis demonstrated that the CS-SAGE framework transforms images into sharply focused and accurately localized results for both point and extended targets. This research addressed limitations in SAR algorithms under multipath conditions.

\bibliographystyle{IEEEtran}
\bibliography{reference}
\end{document}